\documentclass{article}

\usepackage[nonatbib,final,preprint]{neurips_2020}

\usepackage[utf8]{inputenc} 
\usepackage[T1]{fontenc}    
\usepackage{hyperref}       
\usepackage{url}            
\usepackage{booktabs}       
\usepackage{amsfonts}       
\usepackage{nicefrac}       
\usepackage{microtype}      
\usepackage{graphicx}

\usepackage[russian, english]{babel}
\usepackage{CJKutf8}

\title{Automatic Detection and Classification of Tick-borne Skin Lesions using Deep Learning}

\author{%
    Lauren Michelle Pfeifer\\
    Independent Researcher\\
    San Francisco, CA 94107 \\
    \texttt{pfeifer.lauren@gmail.com} \\
    \And
    Matias Valdenegro-Toro \\
    German Research Center for Artificial Intelligence\\
    28359 Bremen, Germany\\
    \texttt{matias.valdenegro@dfki.de} \\
}

\begin{document}
    
    \maketitle
    
    \begin{abstract}
        Around the globe, ticks are the culprit of transmitting a variety of bacterial, viral and parasitic diseases. The incidence of tick-borne diseases has drastically increased within the last decade, with annual cases of Lyme disease soaring to an estimated 300,000 in the United States alone. As a result, more efforts in improving lesion identification approaches and diagnostics for tick-borne illnesses is critical.
        The objective for this study is to build upon the approach used by Burlina et al. by using a variety of convolutional neural network models to detect tick-borne skin lesions. We expanded the data inputs by acquiring images from Google in seven different languages to test if this would diversify training data and improve the accuracy of skin lesion detection. The final dataset included nearly 6,080 images and was trained on a combination of architectures (ResNet 34, ResNet 50, VGG 19, and Dense Net 121). We obtained an accuracy of 80.72\% with our model trained on the DenseNet 121 architecture.
    \end{abstract}
    
    \section{Introduction}
    
    Around the globe, ticks are the culprit of transmitting a variety of bacterial, viral and parasitic diseases  (Rodriguez-Morales, et al., 2018). The incidence of tick-borne diseases, like Lyme disease, Tick-borne Encephalitis, and Rickettsiosis has drastically increased within the last decade \cite{wikel2018ticks} \cite{cdc2019}. The number of annual cases of Lyme disease alone has soared to an estimated 300,000 in the United States \cite{niaid2020}. As a result, more efforts in improving lesion identification approaches and diagnostics for tick-borne illnesses is critical in preventing serious long-term complications \cite{rodriguez2018epidemiology}. 
    
    The initial phase of our research specifically focuses on tick-borne skin lesions relating to Lyme disease, namely erythema migrans (EM), an erythematous, annular skin lesion \cite{bryant2000clinical}. Erythema migrans emerges at the site of the tick bite as an expanding skin redness and often heals without antibiotic treatment, however the infecting pathogens can spread to tissues and organs throughout the body \cite{stanek2018lyme}. Timely diagnosis and treatment of a tick-borne skin lesion, such as EM, can reduce the severity of tick-borne illness symptoms early on \cite{burlina2019automated}. For Lyme disease specifically, deep learning detection models have been applied to skin lesion (erythema migrans) images for the purpose of early diagnostics and treatment. 
    
    \section{Methods and Experimental Results}
    
    Our study is conducting a binary classification of erythema migrans and non-erythema migrans. For non-erythema migrans images, we focused on images of mosquito bites and healthy skin. Mosquito bite images were selected due to the similar appearance in colour and texture (although different in size) from the typical “bulls-eye” shape of erythema migrans. People often mistake erythema migrans as infected mosquito bites \cite{mayo2019} which causes them to forgo getting proper treatment \cite{burlina2019automated}. Similar to the study of Burlina et al., binary classification is our end goal, however this paper focuses on a 3-class classification approach of detecting erythema migrans, mosquito bite(s), and healthy skin.
    
    
    As of now, there are still no publicly available datasets of tick bite images. For this study, the dataset was compiled using images from Google. This approach was leveraged from fastai’s Jeremy Howard and Adrian Rosebrock who use Google Images to gather training data \cite{howard2019} \cite{rosebrock2020}. Our objective is to train a classifier for three classes: tick bites, mosquito bites, and healthy skin, the latters which considers skin without any kind of tick or mosquito bite. 
    
    Our initial search was originally conducted only in English using the search queries of "tick bite", "mosquito bite", and "healthy skin” in Google Images. The results did not populate a diverse outcome of images in terms of skin color, textures, and skin features. For this reason we decided to expand our image data search to include six other languages: Spanish, French, German, Portuguese, Mandarin, and Russian. The selection process for these languages included verifying geographies abroad that have prolific cases of tick-borne diseases, which includes North America, Europe, Northern Asia, and Latin America. We verified that the translations of each search query was correct by using Google Translate. Our final dataset inluded 6,080 images. 
    
    We decided to keep our search queries less complex as a starting point to gather as much information on tick bites as possible. In contrast, Burlina et al. used a combination of search terms (e.g. “Erythema migrans”, “Lyme”, or “bullseye rash”) that are more relevant to Lyme disease \cite{burlina2019automated}. In the next phase of research we plan to fine tune our search queries based on the results we saw from our current search results. Table \ref{query_overview} shows an overview of the queries we used for each language.
    
    
    
    
    The key performance metric in the study was accuracy in demonstrating the number of predictions our model correctly identified as a tick bite, mosquito bite or healthy skin. Out of the four architectures we ran, our model obtained the best accuracy of 80.72\% trained on the DenseNet 121 architecture as indicated in Table \ref{classification_results}. The second best result we obtained was an accuracy of 79.16\% trained on the ResNed 34 architecture. The remaining two architectures, resulted in lower accuracies with VGG-19 at 78.64\% and ResNet 50 at 53.12\%.
    
    In addition to correctly classifying skin lesions, our results of our dataset showed to indeed be less biased in terms of skin color, textures, and skin features. 
    
    
    Our study of using deep learning to automate detection and classification of tick-borne skin lesions generated promising results. The incorporation of using diverse approaches to gather data, such as our use of queries in multiple languages, exhibited a step in the right direction towards reducing bias in skin colors. For instance, the search queries in Portuguese and Spanish for “tick bite” and “mosquito bite” resulted in images on darker skin tones, while the search query for “healthy skin” included more image results with darker skin tones as well. 
    
    As expected, we encountered some limitations from our broad scope of search queries. The results for healthy skin exhibited more challenges in classifying healthy skin. For our next phase of research, we aim to refine the search queries for healthy skin. For tick-borne skin lesions, we also plan to expand the search queries and languages used to cover additional geographies and skin lesions that appear different from erythema migrans, like African Tick Bite Fever or Rickettsiosis. 
    
    \section{Conclusions and Future Work}
    
    
    The results of our model are a great starting point, though we aim for our results to encourage further research in the space of skin-lesion detection for tick-borne illness diagnostics. The accuracy rates exhibit room for improvement before being deployed into medical use. As indicated by Rodriguez-Morales et al., more research, more diagnostics, more surveillance with better identification approaches are very much needed. The final dataset included 6,080 images and was trained on a combination of architectures (ResNet 34, ResNet 50, VGG 19, and Dense Net 121). We obtained an accuracy of 80.72\% with our model trained on the DenseNet 121 architecture.
    
    For next steps, we plan to share the dataset publicly, test out different models such as a classical model approach using a Support Vector Machine on the image pixels. We may also consider testing an ensemble approach, as well as training on Inceptionv4, MobileNet, and EfficientNet to observe whether these models obtain similar accuracies.
    
    \clearpage
    
    \bibliographystyle{plain}
    \bibliography{biblio.bib}
    \clearpage        
    \appendix    
    
    \begin{table}[h]
        \section{Overview of sample images by search query per language}
        \centering
        \begin{tabular}{lllll}
            \toprule
            Language 	& Tick Bite & Mosquito Bite & Healthy Skin 	& Total \\
            \midrule
            English 	& 320 		& 240 			& 320 			& 880 \\
            Spanish 	& 320 		& 320			& 320 			& 960 \\        French 		& 240 		& 320 			& 320 			& 880 \\
            German 		& 240 		& 320 			& 320 			& 880 \\
            Portuguese 	& 240 		& 320 			& 240 			& 800 \\
            Mandarin 	& 320 		& 80 			& 320 			& 720 \\
            Russian 	& 320 		& 320 			& 320 			& 960 \\
            \midrule
            Total		& 2000		& 1920			& 2160			& 6080 \\
            \bottomrule
        \end{tabular}
        \label{dataset_overview}
    \end{table} 
    
    \begin{table}[h]
        \section{Overview of queries for Google Image search in seven selected languages}
        \centering
        \begin{tabular}{llll}
            \toprule
            Language 	& Tick Bite Query & Mosquito Bite Query & Healthy Skin Query \\
            \midrule
            English 	& Tick bite 	  & Mosquito bite 		& Healthy skin 			\\
            Spanish 	& Picadura de garrapata 		& Picadura de mosquito			& Piel saludable 			\\
            French 		& Morsure de tique		& Piqure de Moustique 			& Peau saine 			\\
            German 		& Zeckenbiss 		& Mückenstich 			& Gesunde haut 			\\
            Portuguese 	& Mordida de carrapato 		& Picada de mosquito 			& Pele saudável 			\\
            Mandarin 	& \begin{CJK}{UTF8}{gbsn}蜱咬\end{CJK} &
            \begin{CJK}{UTF8}{gbsn}蚊虫叮咬\end{CJK} &
            \begin{CJK}{UTF8}{gbsn}健康的皮肤\end{CJK}  			\\
            & Pí yǎo		& Wénchóng dīngyǎo		& Jiànkāng de pífū	\\
            Russian 	& \foreignlanguage{russian}{укус клеща} & \foreignlanguage{russian}{комариный укус} & \foreignlanguage{russian}{здоровая кожа} \\
            & ukus kleshcha		& komarinyy ukus	& zdorovaya kozha\\
            \bottomrule
        \end{tabular}
        \label{query_overview}
    \end{table}
    
    \section{Classification Results over multiple neural network architectures}
    
    \begin{table}[h]
        
        \centering
        \begin{tabular}{llll}
            \toprule
            Model	 		& Training Loss & Validation Loss & Validation Accuracy \\
            \midrule
            ResNet-34 		& $0.400625$ 		& $0.702235$  & $79.16 \%$\\
            ResNet-50 		& $0.749036$ 		& $1.147042$  & $53.12 \%$\\
            VGG-19 			& $0.452295$ 		& $0.519739$  & $78.64 \%$\\
            DenseNet-121	& $0.339066$ 		& $0.568043$  & $80.72 \%$\\
            \bottomrule
        \end{tabular}
        \caption{Classification Results over multiple neural network architectures}
        \label{classification_results}
    \end{table}
    
    \clearpage
    \section{Results of search queries run on VGG 19 architecture}
    
    \begin{figure}[h]
        
        \centering
        \includegraphics[width=\textwidth]{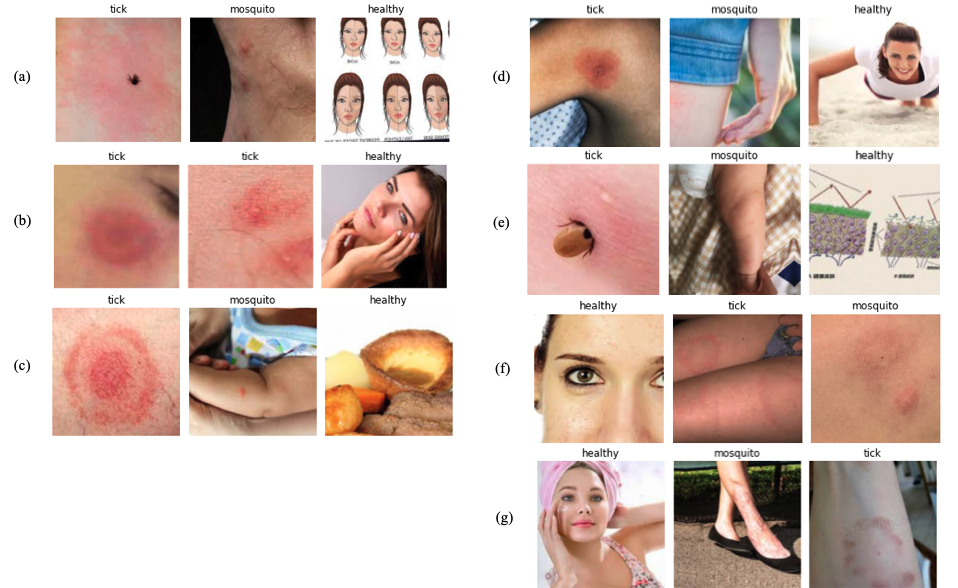}
        \caption{Results of search queries run on VGG 19 architecture in (a) English, (b) Spanish, (c) French, (d) German (e) Madarin (f) Portuguese and (g) Russian.}
        \label{sample_images}
    \end{figure}

    \section{Additional Dataset Images}
    
    \begin{figure}[h]
        \centering
        \includegraphics[width=\textwidth]{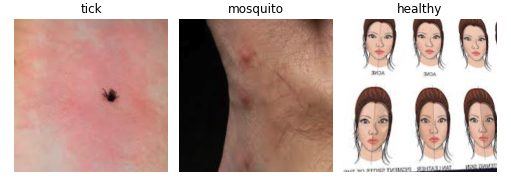}
        \caption{Results from search queries in English}
    \end{figure}
    
    \begin{figure}[hb]
        \centering
        \includegraphics[width=\textwidth]{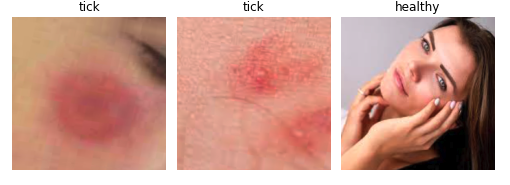}
        \caption{Results from search queries in Spanish}
    \end{figure}
    
    \begin{figure}[hb]
        \centering
        \includegraphics[width=\textwidth]{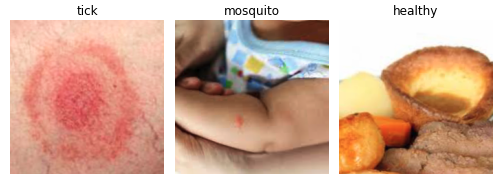}
        \caption{Results from search queries in French}
    \end{figure}
    
    \begin{figure}[hb]
        \centering
        \includegraphics[width=\textwidth]{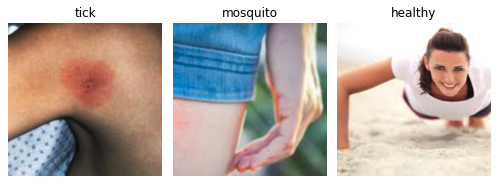}
        \caption{Results from search queries in German}
    \end{figure}
    
    \begin{figure}[hb]
        \centering
        \includegraphics[width=\textwidth]{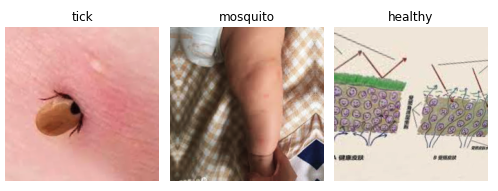}
        \caption{Results from search queries in Mandarin}
    \end{figure}
    
    \begin{figure}[hb]
        \centering
        \includegraphics[width=\textwidth]{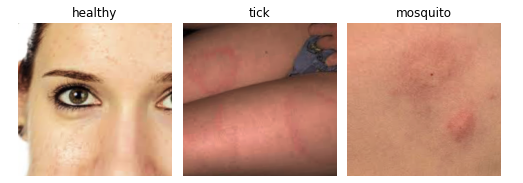}
        \caption{Results from search queries in Portuguese}
    \end{figure}
    
    \begin{figure}[hb]
        \centering
        \includegraphics[width=\textwidth]{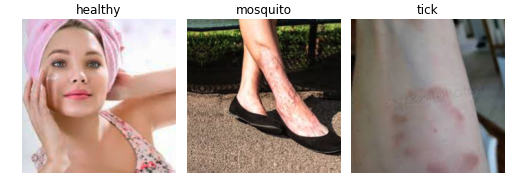}
        \caption{Results from search queries in Russian}
    \end{figure}

\end{document}